\documentclass[english,prl, twocolumn,superscriptaddress,showpacs]{revtex4-1}
\pdfoutput=1
\usepackage[latin1]{inputenc}
\usepackage{verbatim}
\usepackage{amstext}
\usepackage{graphicx}
\usepackage{amssymb}

\makeatletter
\@ifundefined{textcolor}{}
{%
 \definecolor{BLACK}{gray}{0}
 \definecolor{WHITE}{gray}{1}
 \definecolor{RED}{rgb}{1,0,0}
 \definecolor{GREEN}{rgb}{0,1,0}
 \definecolor{BLUE}{rgb}{0,0,1}
 \definecolor{CYAN}{cmyk}{1,0,0,0}
 \definecolor{MAGENTA}{cmyk}{0,1,0,0}
 \definecolor{YELLOW}{cmyk}{0,0,1,0}
 }

\usepackage{leftidx}

\makeatother

\usepackage{babel}

\begin{document}

\title{Decoherence in a Double-Dot Aharonov-Bohm Interferometer}

\author{Bj\"{o}rn Kubala}
\affiliation{Institute for Theoretical Physics, Universit\"{a}t Erlangen-N\"{u}rnberg, Staudtstr. 7, 91058 Erlangen, Germany}
\author{David Roosen}
\affiliation{Institut f\"{u}r Theoretische Physik, Goethe Universit\"{a}t, 60438 Frankfurt/Main, Germany}
\author{Michael Sindel}
\affiliation{Physics Department, ASC and CeNs, Ludwig-Maximilians-Universit\"{a}t, 80333 Munich, Germany}
\author{Walter Hofstetter} 
\affiliation{Institut f\"{u}r Theoretische Physik, Goethe Universit\"{a}t, 60438 Frankfurt/Main, Germany}
\author{Florian Marquardt}
\affiliation{Institute for Theoretical Physics, Universit\"{a}t Erlangen-N\"{u}rnberg, Staudtstr. 7, 91058 Erlangen, Germany}

\date{November 15, 2010}

\begin{abstract}
Coherence in electronic interferometers is typically believed to be
restored fully in the limit of small voltages, frequencies and temperatures.
However, it is crucial to check this essentially perturbative argument
by nonperturbative methods. Here, we use the numerical renormalization
group to study ac transport and decoherence in an experimentally realizable
model interferometer, a parallel double quantum dot coupled to a phonon
mode. The model allows to clearly distinguish renormalization effects
from decoherence. We discuss finite frequency transport and confirm
the restoration of coherence in the dc limit. 
\end{abstract}


\maketitle

\paragraph{Introduction.-}

Quantum coherence and its degradation by interaction with the environment
are crucial ingredients in electronic transport through nanostructures.
Generically, perturbation theory suggests that coherence is fully
restored in the limit of small applied bias voltage, ac frequency,
and temperature. In this limit, the particles do not have enough energy
to leave behind a {}``trace'' in the environment. However, it is
highly desirable to check whether such a statement remains valid beyond
perturbation theory. In this work, we propose to employ the numerical
renormalization group (NRG) \cite{Wilson1975} to study this question.
That method can readily be applied to transport through interacting
localized electronic levels, such as quantum dots and molecules. Investigating
decoherence in such systems offers the advantages of (some degree
of) experimental control of the type and strength of decoherence mechanisms
and, on the theoretical side, of a wealth of powerful, advanced methods.

\begin{figure}[b]
\includegraphics[width=0.9\columnwidth]{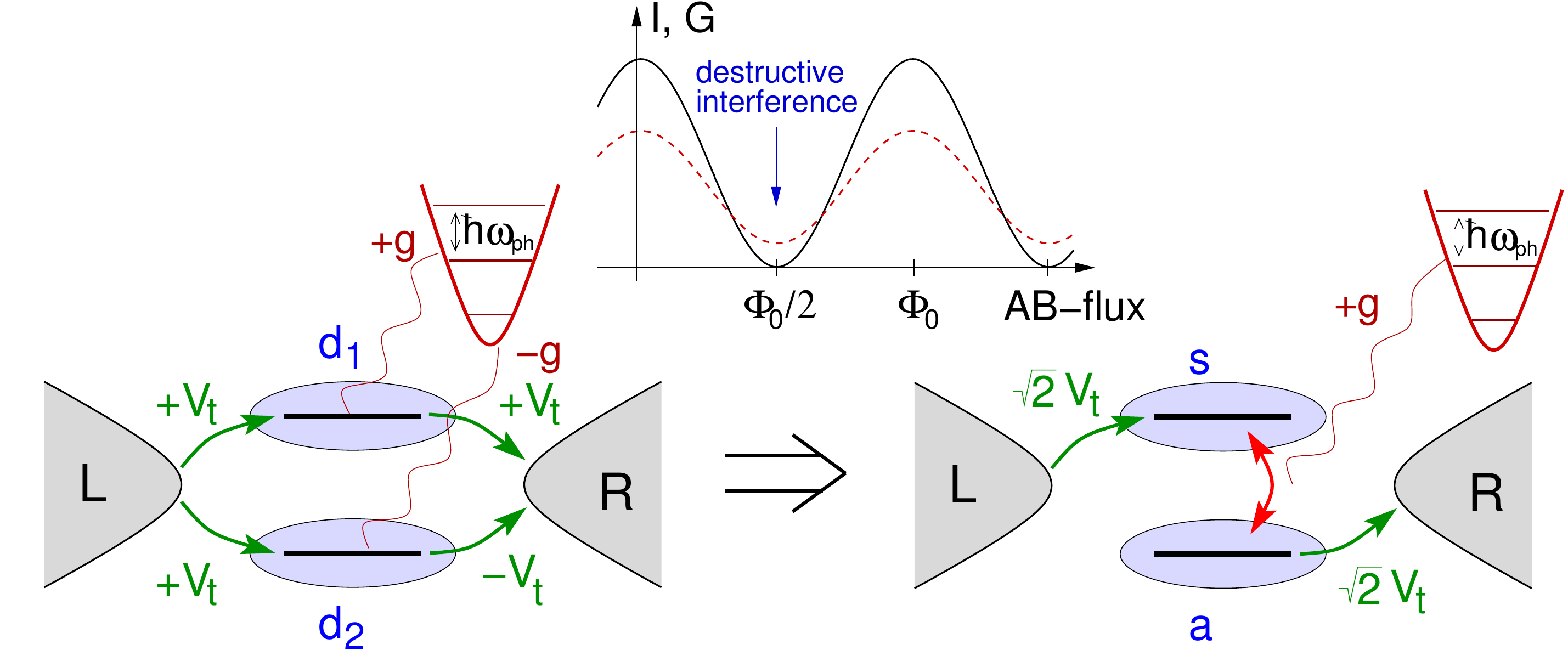}

\caption{\label{fig:Mapping-the-Hamiltonian}A double-dot Aharonov-Bohm interferometer.
Decoherence suppresses the amplitude of AB oscillations (dashed line).
At destructive interference, $\phi_{AB}=\Phi_{0}/2$, only incoherent
transport processes contribute in a symmetric setup. Mapping the Hamiltonian
to (anti)symmetric combinations of the dot levels allows for a transparent
exploitation of this cancellation of coherent transport. }

\end{figure}

In discussing the effects of coupling a system to an environment true
decoherence has to be carefully distinguished from a mere renormalization
of the system's characteristics due to the coupling. Electron-phonon
(e-ph) coupling, for example, affects transport in several ways: Incoherent
processes, in which phonons are absorbed or emitted, may increase
or decrease the total transport through a single site \cite{Tal2008PRL},
and coherent contributions to transport will also be renormalized.
Considering an interferometer - the most obvious way to study decoherence
in electronic transport - it thus turns out to be advantageous
not to consider the overall change in the amplitude of the current
or conductance as a measure of decoherence. Instead, we want to focus
on the special case of destructive interference, i.e., the case of
half a quantum of magnetic flux for an Aharonov-Bohm interferometer
(ABI) (see Fig.~\ref{fig:Mapping-the-Hamiltonian}). The advantage
\cite{Marquardt2003PRB,*Koenig2001PRL} is that in an interferometer with perfectly symmetric arms renormalization
processes will not lift the destructive interference, and any finite
current is a true sign of decoherence.

In this work, we consider a double-dot Aharonov-Bohm interferometer
(see Fig.~\ref{fig:Mapping-the-Hamiltonian}), where electrons on
the dots couple to a single phononic mode, representing the environment.
This scenario can be experimentally realized in tunneling transport
through two quantum dots placed between two electron reservoirs \cite{Holleitner2001PRL,*Sigrist2006PRL}.
On the one hand, one can then consider specifically engineered phononic
environments, for instance, for quantum dots defined within free standing
nanobeams \cite{Weig2004PRL}, or a quantum dot coupling to a cantilever
\cite{BennettPRL10}; on the other hand, the simple scenario can easily
be generalized to model multiple phonon modes in standard heterostructures
or molecular electronics \cite{Galperin2007JPhysC,*Haertle2008PRB}.
We will focus on the linear ac conductance, which is readily accessible
by NRG and of great current interest in general \cite{Gabelli_Science06,Buttiker1993PLA}.
The nonperturbative NRG results will be complemented by nonequilibrium
Green's functions (NEGF) \cite{Rammer1986Quantum} methods.

\paragraph{The Model.-}

Our system is described by the Hamiltonian $\hat{H}=\hat{H}_{\text{imp}}+\sum_{\alpha=L/R}(\hat{H}_{\textrm{res},\alpha}+\hat{H}_{T,\alpha})\,$
with lead Hamiltonian $\hat{H}_{\text{res,}\alpha}=\sum_{k}\epsilon_{k\alpha}\hat{c}_{k\alpha}^{\dagger}\hat{c}_{k\alpha}$,
and spinless electrons on two degenerate dot levels, $n=1,\,2$ coupling
to a single phonon mode, \begin{equation}
\hat{H}_{\textrm{imp}}=\!\!\!\sum_{n=1,2}\!\!\varepsilon\hat{d}_{n}^{\dagger}\hat{d}_{n}+\hbar\omega_{\mathrm{ph}}\hat{b}^{\dagger}\hat{b}+g(\hat{b}+\hat{b}^{\dagger})(\hat{d}_{1}^{\dagger}\hat{d}_{1}-\hat{d}_{2}^{\dagger}\hat{d}_{2})\,,\label{eq:original Hamiltonian}\end{equation}
where the level energy, $\epsilon$, is taken with respect to the
Fermi level of the leads. Assuming an electron-phonon coupling of
equal magnitude, but different signs for the two dots, the phonons
effectively act as a kind of 'which-path' detector decohering the
electrons.

In a symmetric, destructive interferometer, i.e. for a tunneling Hamiltonian,
$\hat{H}_{T,\alpha}=\sum_{k,\, n=1,2}V_{\alpha,n}\hat{c}_{k\alpha}^{\dagger}\hat{d}_{n}+\, c.c.\,,$
where $V_{L/R,\,1}=V_{L,\,2}=-V_{R,\,2}=V_{t}$ it is decoherence
alone, which allows for transport, overcoming the destructive interference.
This is seen by mapping the Hamiltonian to a new basis, $\hat{a}/\hat{s}=(\hat{d}_{1}\mp\hat{d}_{2})/\sqrt{2},$
of (anti)symmetric combinations of the original dot levels. The impurity
Hamiltonian then reads \begin{equation}
\hat{H}_{\mathrm{imp}}=\varepsilon(\hat{s}^{\dagger}\hat{s}+\hat{a}^{\dagger}\hat{a})+g(\hat{b}+\hat{b}^{\dagger})(\hat{s}^{\dagger}\hat{a}+\hat{a}^{\dagger}\hat{s})\label{eq:mapped Hamiltonian}\end{equation}
 with the (anti)symmetric level coupling to the left (right) lead
only with tunneling matrix elements $V_{L,\, s}=V_{R,\, a}=\sqrt{2}V_{t}$
and $V_{R,\, s}=V_{L,\, a}=0$, see Fig.~\ref{fig:Mapping-the-Hamiltonian}.
It decouples into separated left and right parts for $g=0$, whereas
transport across the interferometer becomes possible due to a tunneling
term between the two new levels, which is coupled to the phonon mode.
While it is obvious that the noninteracting contribution vanishes,
we will later show that destructive interference also cancels renormalized
processes, reflected in the vanishing of certain classes of diagrams
after performing the mapping. This particularly advantageous feature
of only observing (in calculation and experiment) incoherent transport
processes offered by the setup considered here was not present in
other theoretical studies of various interference effects in related
setups \cite{Haule1999PRB,*Meden2006PRL,*Hod2006PRL,*Ueda2007NJP,*Gong2008pss} nor
in studies of e-ph interaction for a single dot \cite{Hyldgaard1994AoP,Mitra2004PRB,Egger2008PRB,*Wohlman2009PRB,Wingreen1989PRB,*Flensberg2003PRB,*Koch2004PRB}.

\paragraph{Techniques.-}

We will present two approaches to study this system. \emph{First},
we employ non-equilibrium Green's function (NEGF) theory, where, based
on the exact noninteracting result \cite{Kubala2002PRB,*Silva2002PRB},
electron-phonon interaction is perturbatively included using Keldysh
diagrams. Particularly tailored to treat nonequilibrium situations,
which can easily be realized in quantum dot systems by an applied
dc bias, the NEGF or Keldysh method \cite{Rammer1986Quantum} has
been widely used as a physically transparent tool to investigate e-ph
interaction in such systems, often, as here, in a perturbative manner
\cite{Hyldgaard1994AoP,Mitra2004PRB,Egger2008PRB,*Wohlman2009PRB}
or employing extensions of the polaron approach \cite{Wingreen1989PRB,*Flensberg2003PRB,*Koch2004PRB,Mitra2004PRB},
which allows tackling strong electron-phonon coupling for weakly tunnel-coupled
systems. In this work, we focus on the \emph{linear ac-conductance} $\mathcal{G}_{\alpha\beta}$, giving
the finite frequency current $I_{\alpha}(\omega_{\textrm{ac}})=\mathfrak{\mathcal{G}}_{\alpha\beta}(\omega_{\textrm{ac}})\, V_{\beta}(\omega_{\textrm{ac}})$
flowing out of lead $\alpha$ as response to a small ac-excitation
voltage $V_{\beta}$ of frequency $\omega_{\textrm{ac}}$ applied
to lead $\beta$ %
\footnote{Here, as in the remainder of this paper, we consider particle currents.
The displacement-current contribution has to be treated separately
\cite{Buttiker1993PLA}.%
}. It is connected to a (retarded) current-current correlator, $\mathcal{G}_{\alpha\beta}(\omega_{\textrm{ac}})=\left[K_{\alpha\beta}(\omega_{\textrm{ac}})-K_{\alpha\beta}(0)\right]/(i\omega_{\textrm{ac}})\,$,
where\begin{equation}
K_{\alpha\beta}(\omega_{\textrm{ac}})=-\frac{i}{\hbar}\int_{0}^{\infty}dt\; e^{i\omega_{\textrm{ac}}t}\langle\left[\hat{I}_{\alpha}(t),\;\hat{I}_{\beta}(0)\right]\rangle\;.\label{eq:correlator}\end{equation}
To calculate this correlator, and, hence, the ac conductance, we use
a recently developed Meir-Wingreen-type formula for the linear ac
conductance \cite{Kubala2010PRB}, which allows straightforward diagrammatic
calculations perturbatively including electron-phonon interaction
- complementing previous works building on a fully time-dependent
Green's function formalism \cite{JauhoPRB94} (see e.g., \cite{Li1996Quantum},
treating electron-electron interactions) and rate-equation approaches
\cite{Lehmann2004JCP}.

While the perturbative NEGF-approach is only valid for weak electron-phonon
interaction, our \emph{second} approach, the numerical renormalization
group method (NRG), invented by Wilson in the late 1970s \cite{Wilson1975},
is especially tailored for strongly interacting systems. Throughout
the last decades numerous enhancements of the numerical renormalization
group method have been introduced, ever improving the accuracy of
correlation functions \cite{Costi1994,Hofstetter2000,Weichselbaum2007,Bulla2008}
and opening the possibility to study quantum impurity systems coupled
to a bosonic bath \cite{Bulla2003}. Here, we present to our best
knowledge the first NRG calculation of ac conductance in a quantum
dot system with an additional bosonic degree of freedom (cf.~NRG
studies on ac conductance with electron-electron interaction, for
instance, in the Kondo regime \cite{Sindel2005PRL,*Moca2010PRB}, and
investigations of electron-phonon coupling effects on linear dc-conductance,
spectral density and spin and charge susceptibilities in \cite{Hewson2002,Cornaglia2004,*Cornaglia2005}).
For a quantum impurity system with a fermionic bath and an additional
bosonic mode coupled to the impurity directly, the numerical renormalization
group can be applied in a similar fashion as in the {}``pure'' fermionic
case without the bosonic mode. The mapping to the Wilson chain is
not affected by the additional bosonic mode \cite{Bulla2008,Hewson2002},
since the bosons only enter in the very first step of the iterative
diagonalization procedure. We implemented a NRG algorithm using the
reduced density matrix \cite{Hofstetter2000} as well as the complete basis of the underlying Fock space. Furthermore, we performed
a $z$-trick averaging \cite{Yoshida1990,*Zitko2009} over
$32$ slightly different discretizations of the conduction band to
improve on the logarithmic resolution of energy (and thereby also
frequency) around the Fermi energy inherent to NRG. In that manner,
reliable numerical results up to frequencies of some ten percent of
the bandwidth could be achieved with reasonably small numerical errors
and artifacts.%
\begin{figure}
\includegraphics[width=0.9\columnwidth]{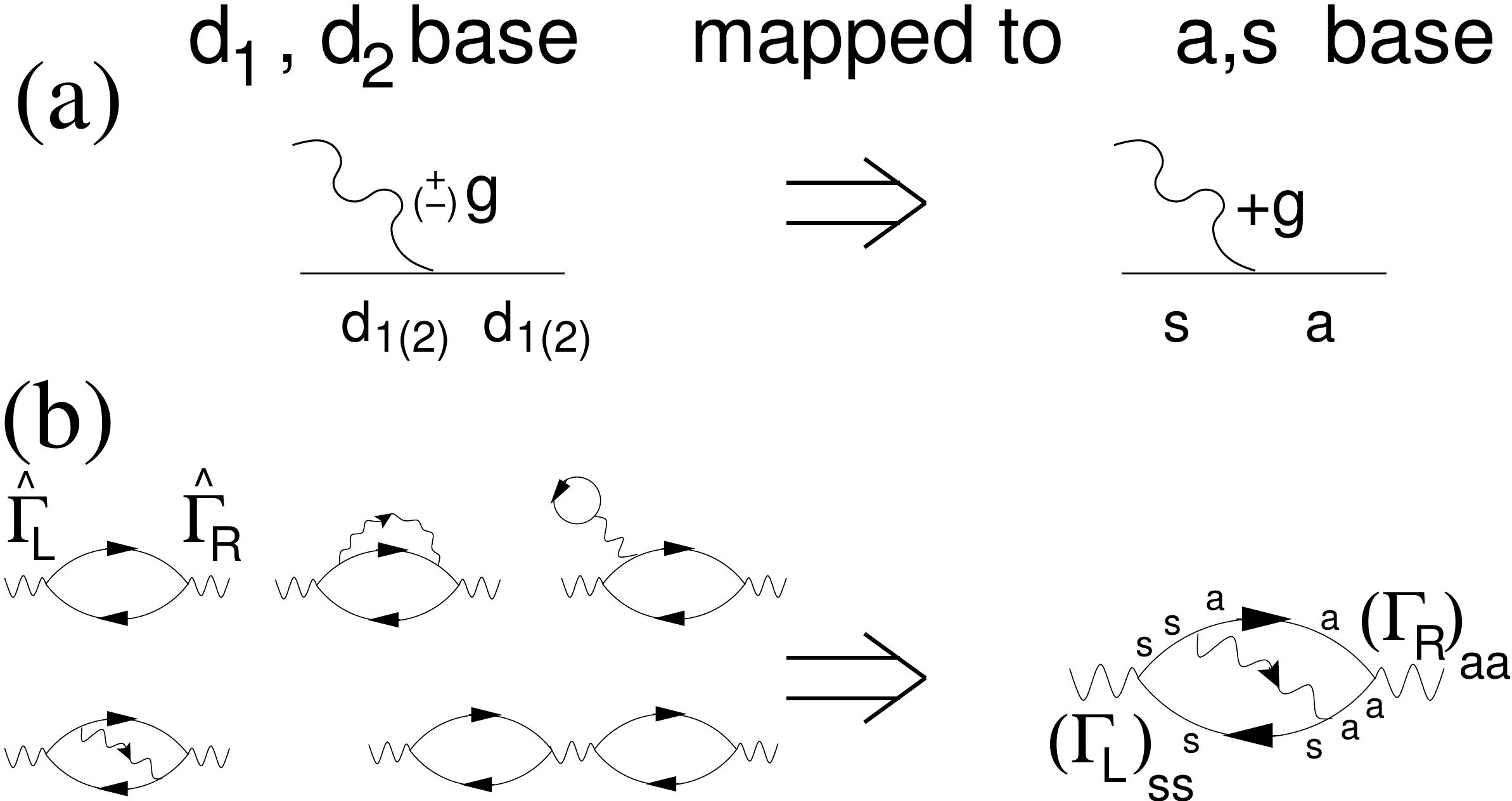}

\caption{\label{fig:Mapping-the-Hamiltonian-diags}Mapping the Hamiltonian
to (anti)symmetric combinations of the dot levels: the mapped electron-phonon
interaction vertex (a) restricts the number and type of diagrams in
a perturbative calculation of ac conductance (b), reflecting the destructive
interference of all coherent contributions. (Solid lines denote electron
Green's functions, wiggly lines stand for phonon propagators.)}

\end{figure}

\paragraph{Results and interpretation.-}

\begin{figure}
\includegraphics[bb=20bp 40bp 708bp 500bp,width=1\columnwidth]{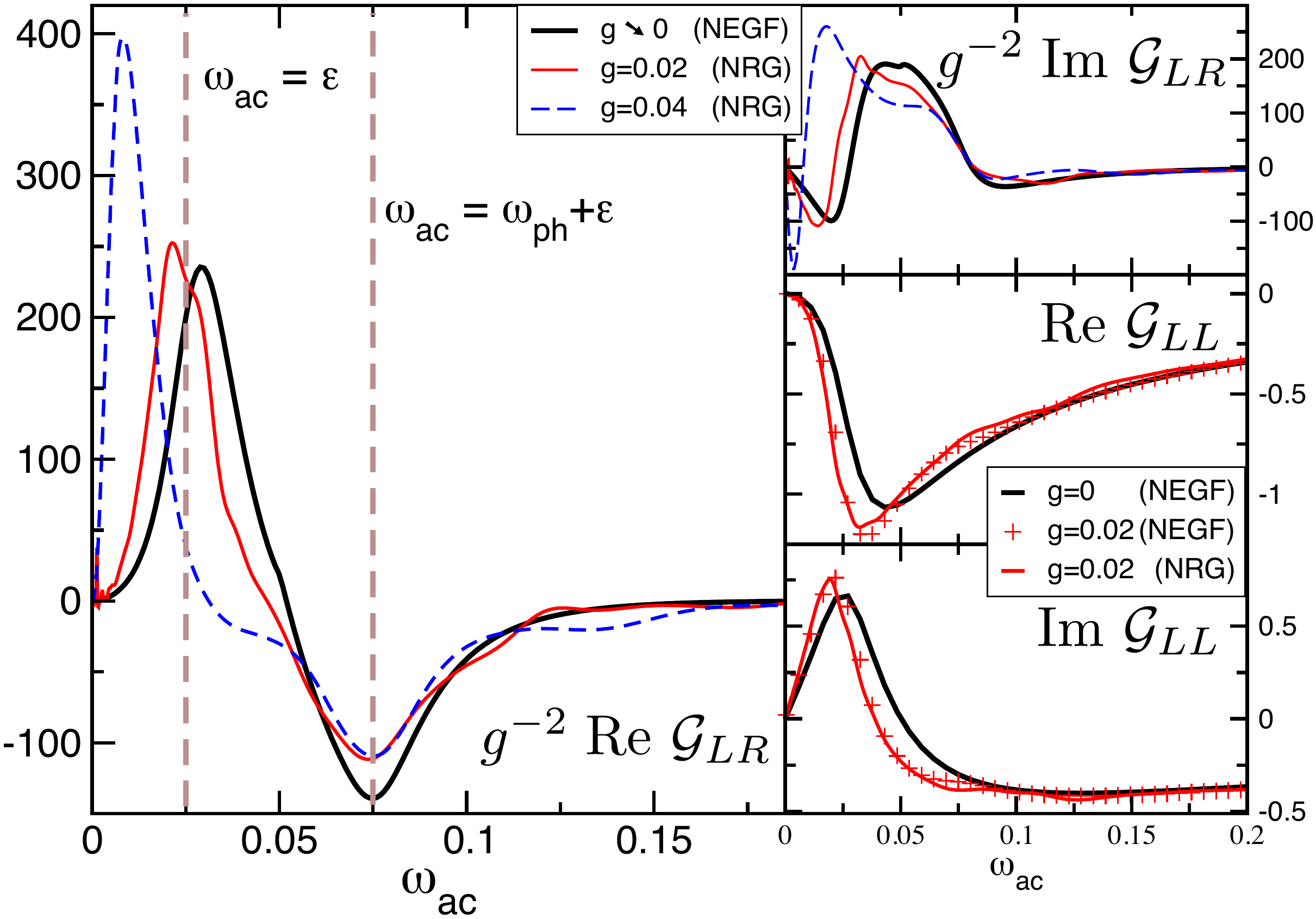}\caption{\label{fig:Strong-electron-phonon-coupling}ac conductance through
the destructive ABI. In the limit of weak electron-phonon coupling
NRG and perturbative calculations match. For stronger e-ph coupling,
$g/\omega_{\mathrm{ph}}\approx1$, additional sidepeaks
at $\omega_{\mathrm{ac}}=\epsilon+n\omega_{\mathrm{ph}}\quad(n=1,\,2,\ldots)$
appear, and the first resonance is shifted to lower frequency. $\mathcal{G}_{LR}(\omega_\mathrm{ac}=0)\equiv0$
is found also for strong e-ph coupling. Parameters are $\omega_{\mathrm{ph}}=0.05,\;\epsilon=0.025,\;\Gamma=2\pi(\rho_{L}+\rho_{R})|V_{t}|^{2}=0.02$
in units of the bandwidth, conductances in units of $e^{2}/h$, and
we consider the zero temperature limit and no dc-bias.}

\end{figure}

To investigate the effects of \emph{weak electron-phonon coupling}
on ac-transport, we employ NEGF, considering contributions to the
ac-conductance from diagrams with up to a single phonon line, which
gives contributions up to order ${\cal O}(g^2)$.
Using the mapping to (anti)symmetric dot levels in a diagrammatic
representation for the ac conductance automatically exploits the main
characteristic of transport in the symmetric destructive ABI, namely
the complete cancellation of coherent transport processes due to destructive
interference. For instance, considering the ac conductance, $\mathcal{G}_{LR}(\omega_\mathrm{ac})$,
there are a zeroth-order and several second-order diagrams constructable
from the original e-ph interaction vertex, Fig.~\ref{fig:Mapping-the-Hamiltonian-diags},
while no zeroth-order and only a single second-order diagram can be
constructed from the mapped e-ph coupling vertex and the mapped lead-dot
couplings. The nonexistence of those diagrams reflects the notion
that in the destructive ABI the coherent zero-order contribution,
as well as contributions, where transport is {}``merely renormalized'',
will interfere destructively and thus vanish. There are, however,
diagrammatic contributions for the {}``diagonal'' conductances $\mathcal{G}_{LL/RR}(\omega_\mathrm{ac})$,
describing electrons tunneling back and forth between the (anti)symmetric
level and the (left) right lead, as the chemical potential of the
lead is varied with frequency $\omega_\mathrm{ac}$. The noninteracting result
thus found (see the lower right panels of Fig.~\ref{fig:Strong-electron-phonon-coupling}),
which is also accessible by simpler techniques \cite{Buttiker1993PLA,JauhoPRB94},
reproduces for small frequencies the universal behavior of a mesoscopic
RC-circuit with a quantized charge relaxation resistance \cite{Gabelli_Science06,Buttiker1993PLA}.
The second-order contribution to $\mathcal{G}_{LL}$ constitutes a
small correction to the noninteracting result, which yields additional
features at $\omega_\mathrm{ac}=\epsilon+\omega_{\mathrm{ph}}$. Turning to the
conductance across the interferometer, $\mathcal{G}_{LR}$, we find,
as argued above, no zeroth order, and only a single, topologically
distinct diagram in second order, namely an electronic loop with a
crossing phonon line {[}see Fig.~\ref{fig:Mapping-the-Hamiltonian-diags}(b){]}.
One can now easily use Keldysh diagrammatic rules to find the corresponding
integral expression for the ac conductance in terms of the electronic
Green's functions and the bare phononic propagators $D_{E'}^{R/A}=(E'-\omega_{\text{ph}}\pm i\eta)^{-1}-(E'+\omega_{\text{ph}}\pm i\eta)^{-1}$.
Interestingly, there are contributions from the $\delta$-peak at
the phonon energy $\omega_\mathrm{ph}$ as well as from the principal-value
part. For the dissipative real part of the conductance, $\mathrm{Re}\,\mathcal{G}_{LR}$,
(see left panel of Fig.~\ref{fig:Strong-electron-phonon-coupling}),
we find that contributions from the latter yield a resonant peak at
$\omega_\mathrm{ac}=\epsilon$. Around $\omega_\mathrm{ac}=\epsilon+\omega_\mathrm{ph}$,
where the excitation frequency is in resonance with the dot level
position after the excitation of a phonon, there is a smaller negative
peak, stemming from the $\delta$-contributions. Corresponding features
at the same frequencies are found for the imaginary part, $\mathrm{Im}\,\mathcal{G}_{LR}$.

Investigating\emph{ strong electron-phonon coupling}, comparisons
of NRG calculations to the perturbative results are shown in Fig.~\ref{fig:Strong-electron-phonon-coupling}.
For weak e-ph coupling, there is good qualitative and also quantitative
agreement between NRG and the perturbative results: In the limit $g\searrow0$,
we reach the noninteracting results for $\mathcal{G}_{LL}$(thick
black lines in the two lower right panels of Fig.~\ref{fig:Strong-electron-phonon-coupling}),
while for $\mathcal{G}_{LR}$ we recover the perturbative, ${\cal O}(g^{2})$,
result derived above. The left panel of Fig.~\ref{fig:Strong-electron-phonon-coupling}
also demonstrates the main effects of stronger e-ph coupling on $\mathrm{Re}\,\mathcal{G}_{LR}$.
Most importantly, one finds additional resonances at sidebands of
the level energy, $\omega_{\mathrm{ac}}=\epsilon+n\omega_{\mathrm{ph}}\enskip(n=1,\,2,\ldots)$,
which stem from processes involving several phonons. Furthermore, we observe a shift of the first resonance from $\omega_\mathrm{ac}\approx\epsilon$
to smaller frequencies reminiscent of a polaron shift.

In the following, we want to highlight two particular remarkable features
of our results at low frequency. The \emph{first}, observable already
in the weak coupling limit, is the fact, that ac transport is already
possible for excitation frequencies smaller than the phonon energy,
$\omega_{\mathrm{ac}}<\omega_{\mathrm{ph}}\,$. Let us first emphasize,
that this is in striking contrast to the case of finite-bias dc transport,
where in the weak coupling limit there is a sharp threshold for incoherent
transport processes becoming possible only for $eV_{\mathrm{dc}}\ge\omega_{\mathrm{ph}}$.
Using the mapping to (anti)symmetric dot states the dc current through
a destructive ABI can be shown to contain only incoherent contributions.
These contributions are characterized by explicit appearance (cf.
Eq.~66 in Ref.~\cite{Hyldgaard1994AoP}) of self-energies, $\Sigma^{\gtrless,\,\mathrm{ph}}$,
due to emission or absorption of phonons. In lowest order in the e-ph
coupling (and for zero temperature) transport takes place through
two {}``transport channels'' at $\omega=\epsilon,\;\epsilon+\omega_{\mathrm{ph}}\,$
within a transport window, $\mu_{L}>\omega>\mu_{R}+\omega_{\mathrm{ph}}\,$.
In particular, the resulting I-V curve has a sharp gap at low bias,
$eV_{\mathrm{dc}}<\omega_{\mathrm{ph}}$, as any incoherent transport
process necessarily involves the emission of a phonon, whose energy
has to be provided by the applied bias. The naive expectation, that
in ac transport the frequency of the infinitesimal ac-excitation voltage
providing energy for emission of phonons acts in a similar manner
does not hold. In fact, subthreshold ac transport apparently is possible
without the real emission of phonons. Considering the Hamiltonian
of the destructive ABI in the mapped base, we see that virtual tunneling
back and forth processes between symmetric and antisymmetric levels
with emission and absorption of phonons will yield an effective charge-charge
interaction between the two sites. Such interaction then leads to
finite ac conductance, but vanishing dc conductance, just like the
electrostatic interaction in a plate capacitance.

While both of these arguments explain the vanishing of conductance
in the dc limit in lowest order in the e-ph coupling, note as a \emph{second}, most important feature of the results in Fig.~\ref{fig:Strong-electron-phonon-coupling}, that our NRG results
indicate that this holds for strong coupling also. Indeed, while in
a lowest-order process phonons can only be excited exactly at their
bare frequency $\omega_{\mathrm{ph}}\,$, considering higher orders,
one may naturally ask about the effects of the broadening of the phonon
propagator due to coupling to the electrons. Some (though not all)
higher-order effects can thus be captured by calculating the electronic
self-energies with a phonon propagator dressed with electronic polarization
bubbles. In the dc-case, one finds that electrons can indeed absorb
an energy up to $eV$ from the phononic bath broadened by coupling
to the electrons, instead of needing to emit $\omega_{\mathrm{ph}}$
as in lowest order. Nonetheless, the weight of such contributions
is small leading to a current dependence $I\propto(eV)^{3}$ for small
bias voltage for the considered higher-order contributions in the
e-ph coupling and consequently a vanishing linear conductance at zero
frequency. This agrees with the dc limit of our NRG results for the
ac conductance, which indicates as discussed that all higher-order
contributions to the linear zero-temperature conductance vanish. We
thus confirm the naive perturbative argument, that there are no incoherent transport
processes at vanishing frequency, temperature and transport voltage.

\paragraph{Conclusions.-}

We have studied decoherence and the effects of electron-phonon coupling
on ac transport in a double-dot interferometer. A diagrammatic calculation
reveals that renormalization processes can be clearly distinguished
from decoherence for the case of destructive interference, and it
explains low frequency transport as a consequence of phonon-mediated
charge-charge interactions in a mapped system. NRG calculations for
stronger coupling show characteristic features due to multi-phonon
processes. With increasing coupling, the phonon mode broadens and
effectively yields a continuous environment spectrum. Nevertheless,
the nonperturbative NRG result confirms that coherence is fully restored
in the zero-frequency limit of the linear conductance at zero temperature.

We acknowledge support by the DFG (Emmy-Noether, NIM, SFB TR 12),
GIF and DIP.

\bibliographystyle{apsrev}
\bibliography{AB_paper_from_ac_paper}

\begin{thebibliography}{42}
\expandafter\ifx\csname natexlab\endcsname\relax\def\natexlab#1{#1}\fi
\expandafter\ifx\csname bibnamefont\endcsname\relax
  \def\bibnamefont#1{#1}\fi
\expandafter\ifx\csname bibfnamefont\endcsname\relax
  \def\bibfnamefont#1{#1}\fi
\expandafter\ifx\csname citenamefont\endcsname\relax
  \def\citenamefont#1{#1}\fi
\expandafter\ifx\csname url\endcsname\relax
  \def\url#1{\texttt{#1}}\fi
\expandafter\ifx\csname urlprefix\endcsname\relax\def\urlprefix{URL }\fi
\providecommand{\bibinfo}[2]{#2}
\providecommand{\eprint}[2][]{\url{#2}}

\bibitem[{\citenamefont{Wilson}(1975)}]{Wilson1975}
\bibinfo{author}{\bibfnamefont{K.~G.} \bibnamefont{Wilson}},
  \bibinfo{journal}{\rmp} \textbf{\bibinfo{volume}{47}}, \bibinfo{pages}{773}
  (\bibinfo{year}{1975}).

\bibitem[{\citenamefont{Tal et~al.}(2008)\citenamefont{Tal, Krieger, Leerink,
  and van Ruitenbeek}}]{Tal2008PRL}
\bibinfo{author}{\bibfnamefont{O.}~\bibnamefont{Tal}},
  \bibinfo{author}{\bibfnamefont{M.}~\bibnamefont{Krieger}},
  \bibinfo{author}{\bibfnamefont{B.}~\bibnamefont{Leerink}}, \bibnamefont{and}
  \bibinfo{author}{\bibfnamefont{J.~M.} \bibnamefont{van Ruitenbeek}},
  \bibinfo{journal}{\prl} \textbf{\bibinfo{volume}{100}},
  \bibinfo{pages}{196804} (\bibinfo{year}{2008}) and Refs. within.

\bibitem[{\citenamefont{Marquardt and Bruder}(2003)}]{Marquardt2003PRB}
\bibinfo{author}{\bibfnamefont{F.}~\bibnamefont{Marquardt}} \bibnamefont{and}
  \bibinfo{author}{\bibfnamefont{C.}~\bibnamefont{Bruder}},
  \bibinfo{journal}{\prb} \textbf{\bibinfo{volume}{68}},
  \bibinfo{pages}{195305} (\bibinfo{year}{2003});
\bibitem[{\citenamefont{K\"{o}nig and Gefen}(2001)}]{Koenig2001PRL}
\bibinfo{author}{\bibfnamefont{J.}~\bibnamefont{K\"{o}nig}} \bibnamefont{and}
  \bibinfo{author}{\bibfnamefont{Y.}~\bibnamefont{Gefen}},
  \bibinfo{journal}{\prl} \textbf{\bibinfo{volume}{86}}, \bibinfo{pages}{3855}
  (\bibinfo{year}{2001}).

\bibitem[{\citenamefont{Holleitner et~al.}(2001)\citenamefont{Holleitner,
  Decker, Qin, Eberl, and Blick}}]{Holleitner2001PRL}
\bibinfo{author}{\bibfnamefont{A.~W.} \bibnamefont{Holleitner}},
  \bibinfo{author}{\bibfnamefont{C.~R.} \bibnamefont{Decker}},
  \bibinfo{author}{\bibfnamefont{H.}~\bibnamefont{Qin}},
  \bibinfo{author}{\bibfnamefont{K.}~\bibnamefont{Eberl}}, \bibnamefont{and}
  \bibinfo{author}{\bibfnamefont{R.~H.} \bibnamefont{Blick}},
  \bibinfo{journal}{\prl} \textbf{\bibinfo{volume}{87}},
  \bibinfo{pages}{256802} (\bibinfo{year}{2001});
\bibitem[{\citenamefont{Sigrist et~al.}(2006)\citenamefont{Sigrist, Ihn,
  Ensslin, Loss, Reinwald, and Wegscheider}}]{Sigrist2006PRL}
\bibinfo{author}{\bibfnamefont{M.}~\bibnamefont{Sigrist}},
  \bibinfo{author}{\bibfnamefont{T.}~\bibnamefont{Ihn}},
  \bibinfo{author}{\bibfnamefont{K.}~\bibnamefont{Ensslin}},
  \bibinfo{author}{\bibfnamefont{D.}~\bibnamefont{Loss}},
  \bibinfo{author}{\bibfnamefont{M.}~\bibnamefont{Reinwald}}, \bibnamefont{and}
  \bibinfo{author}{\bibfnamefont{W.}~\bibnamefont{Wegscheider}},
  \bibinfo{journal}{\prl} \textbf{\bibinfo{volume}{96}},
  \bibinfo{pages}{036804} (\bibinfo{year}{2006}).

\bibitem[{\citenamefont{Weig et~al.}(2004)\citenamefont{Weig, Blick, Brandes,
  Kirschbaum, Wegscheider, Bichler, and Kotthaus}}]{Weig2004PRL}
\bibinfo{author}{\bibfnamefont{E.~M.} \bibnamefont{Weig}},
  \bibinfo{author}{\bibfnamefont{R.~H.} \bibnamefont{Blick}},
  \bibinfo{author}{\bibfnamefont{T.}~\bibnamefont{Brandes}},
  \bibinfo{author}{\bibfnamefont{J.}~\bibnamefont{Kirschbaum}},
  \bibinfo{author}{\bibfnamefont{W.}~\bibnamefont{Wegscheider}},
  \bibinfo{author}{\bibfnamefont{M.}~\bibnamefont{Bichler}}, \bibnamefont{and}
  \bibinfo{author}{\bibfnamefont{J.~P.} \bibnamefont{Kotthaus}},
  \bibinfo{journal}{\prl} \textbf{\bibinfo{volume}{92}},
  \bibinfo{pages}{046804} (\bibinfo{year}{2004}).

\bibitem[{\citenamefont{Bennett et~al.}(2010)\citenamefont{Bennett, Cockins,
  Miyahara, Gr\"utter, and Clerk}}]{BennettPRL10}
\bibinfo{author}{\bibfnamefont{S.~D.} \bibnamefont{Bennett}},
  \bibinfo{author}{\bibfnamefont{L.}~\bibnamefont{Cockins}},
  \bibinfo{author}{\bibfnamefont{Y.}~\bibnamefont{Miyahara}},
  \bibinfo{author}{\bibfnamefont{P.}~\bibnamefont{Gr\"utter}},
  \bibnamefont{and} \bibinfo{author}{\bibfnamefont{A.~A.} \bibnamefont{Clerk}},
  \bibinfo{journal}{\prl} \textbf{\bibinfo{volume}{104}},
  \bibinfo{pages}{017203} (\bibinfo{year}{2010}).

\bibitem[{\citenamefont{Galperin et~al.}(2007)\citenamefont{Galperin, Ratner,
  and Nitzan}}]{Galperin2007JPhysC}
\bibinfo{author}{\bibfnamefont{M.}~\bibnamefont{Galperin}},
  \bibinfo{author}{\bibfnamefont{M.~A.} \bibnamefont{Ratner}},
  \bibnamefont{and} \bibinfo{author}{\bibfnamefont{A.}~\bibnamefont{Nitzan}},
  \bibinfo{journal}{J. Phys. Condens. Matter} \textbf{\bibinfo{volume}{19}},
  \bibinfo{pages}{103201} (\bibinfo{year}{2007});
\bibitem[{\citenamefont{H\"{a}rtle et~al.}(2008)\citenamefont{H\"{a}rtle,
  Benesch, and Thoss}}]{Haertle2008PRB}
\bibinfo{author}{\bibfnamefont{R.}~\bibnamefont{H\"{a}rtle}},
  \bibinfo{author}{\bibfnamefont{C.}~\bibnamefont{Benesch}}, \bibnamefont{and}
  \bibinfo{author}{\bibfnamefont{M.}~\bibnamefont{Thoss}},
  \bibinfo{journal}{\prb} \textbf{\bibinfo{volume}{77}},
  \bibinfo{pages}{205314} (\bibinfo{year}{2008}).

\bibitem[{\citenamefont{Gabelli et~al.}(2006)\citenamefont{Gabelli, Feve,
  Berroir, Placais, Cavanna, Etienne, Jin, and Glattli}}]{Gabelli_Science06}
\bibinfo{author}{\bibfnamefont{J.}~\bibnamefont{Gabelli}},
  \bibinfo{author}{\bibfnamefont{G.}~\bibnamefont{Feve}},
  \bibinfo{author}{\bibfnamefont{J.~M.} \bibnamefont{Berroir}},
  \bibinfo{author}{\bibfnamefont{B.}~\bibnamefont{Placais}},
  \bibinfo{author}{\bibfnamefont{A.}~\bibnamefont{Cavanna}},
  \bibinfo{author}{\bibfnamefont{B.}~\bibnamefont{Etienne}},
  \bibinfo{author}{\bibfnamefont{Y.}~\bibnamefont{Jin}}, \bibnamefont{and}
  \bibinfo{author}{\bibfnamefont{D.~C.} \bibnamefont{Glattli}},
  \bibinfo{journal}{Science} \textbf{\bibinfo{volume}{313}},
  \bibinfo{pages}{499} (\bibinfo{year}{2006}).

\bibitem[{\citenamefont{B\"{u}ttiker et~al.}(1993)\citenamefont{B\"{u}ttiker,
  Thomas, and Pr\^{e}tre}}]{Buttiker1993PLA}
\bibinfo{author}{\bibfnamefont{M.}~\bibnamefont{B\"{u}ttiker}},
  \bibinfo{author}{\bibfnamefont{H.}~\bibnamefont{Thomas}}, \bibnamefont{and}
  \bibinfo{author}{\bibfnamefont{A.}~\bibnamefont{Pr\^{e}tre}},
  \bibinfo{journal}{Phys. Lett. A} \textbf{\bibinfo{volume}{180}},
  \bibinfo{pages}{364 } (\bibinfo{year}{1993}).

\bibitem[{\citenamefont{Rammer and Smith}(1986)}]{Rammer1986Quantum}
\bibinfo{author}{\bibfnamefont{J.}~\bibnamefont{Rammer}} \bibnamefont{and}
  \bibinfo{author}{\bibfnamefont{H.}~\bibnamefont{Smith}},
  \bibinfo{journal}{\rmp} \textbf{\bibinfo{volume}{58}}, \bibinfo{pages}{323}
  (\bibinfo{year}{1986}).

\bibitem[{\citenamefont{Haule and Bon\v{c}a}(1999)}]{Haule1999PRB}
\bibinfo{author}{\bibfnamefont{K.}~\bibnamefont{Haule}} \bibnamefont{and}
  \bibinfo{author}{\bibfnamefont{J.}~\bibnamefont{Bon\v{c}a}},
  \bibinfo{journal}{\prb} \textbf{\bibinfo{volume}{59}}, \bibinfo{pages}{13087}
  (\bibinfo{year}{1999});
\bibitem[{\citenamefont{Meden and Marquardt}(2006)}]{Meden2006PRL}
\bibinfo{author}{\bibfnamefont{V.}~\bibnamefont{Meden}} \bibnamefont{and}
  \bibinfo{author}{\bibfnamefont{F.}~\bibnamefont{Marquardt}},
  \bibinfo{journal}{\prl} \textbf{\bibinfo{volume}{96}},
  \bibinfo{pages}{146801} (\bibinfo{year}{2006});
\bibitem[{\citenamefont{Hod et~al.}(2006)\citenamefont{Hod, Baer, and
  Rabani}}]{Hod2006PRL}
\bibinfo{author}{\bibfnamefont{O.}~\bibnamefont{Hod}},
  \bibinfo{author}{\bibfnamefont{R.}~\bibnamefont{Baer}}, \bibnamefont{and}
  \bibinfo{author}{\bibfnamefont{E.}~\bibnamefont{Rabani}},
  \bibinfo{journal}{\prl} \textbf{\bibinfo{volume}{97}},
  \bibinfo{pages}{266803} (\bibinfo{year}{2006});
\bibitem[{\citenamefont{Ueda and Eto}(2007)}]{Ueda2007NJP}
\bibinfo{author}{\bibfnamefont{A.}~\bibnamefont{Ueda}} \bibnamefont{and}
  \bibinfo{author}{\bibfnamefont{M.}~\bibnamefont{Eto}}, \bibinfo{journal}{New
  J. Phys.} \textbf{\bibinfo{volume}{9}}, \bibinfo{pages}{119}
  (\bibinfo{year}{2007});
\bibitem[{\citenamefont{Gong et~al.}(2008)\citenamefont{Gong, Zheng, Wang, and
  L\"{u}}}]{Gong2008pss}
\bibinfo{author}{\bibfnamefont{W.}~\bibnamefont{Gong}},
  \bibinfo{author}{\bibfnamefont{Y.}~\bibnamefont{Zheng}},
  \bibinfo{author}{\bibfnamefont{J.}~\bibnamefont{Wang}}, \bibnamefont{and}
  \bibinfo{author}{\bibfnamefont{T.}~\bibnamefont{L\"{u}}},
  \bibinfo{journal}{Physs Status Solidi B} \textbf{\bibinfo{volume}{245}},
  \bibinfo{pages}{1175} (\bibinfo{year}{2008}).

\bibitem[{\citenamefont{Hyldgaard et~al.}(1994)\citenamefont{Hyldgaard,
  Hershfield, Davies, and Wilkins}}]{Hyldgaard1994AoP}
\bibinfo{author}{\bibfnamefont{P.}~\bibnamefont{Hyldgaard}},
  \bibinfo{author}{\bibfnamefont{S.}~\bibnamefont{Hershfield}},
  \bibinfo{author}{\bibfnamefont{J.~H.} \bibnamefont{Davies}},
  \bibnamefont{and} \bibinfo{author}{\bibfnamefont{J.~W.}
  \bibnamefont{Wilkins}}, \bibinfo{journal}{Ann. Phys. (NY)}
  \textbf{\bibinfo{volume}{236}}, \bibinfo{pages}{1 } (\bibinfo{year}{1994}).

\bibitem[{\citenamefont{Mitra et~al.}(2004)\citenamefont{Mitra, Aleiner, and
  Millis}}]{Mitra2004PRB}
\bibinfo{author}{\bibfnamefont{A.}~\bibnamefont{Mitra}},
  \bibinfo{author}{\bibfnamefont{I.}~\bibnamefont{Aleiner}}, \bibnamefont{and}
  \bibinfo{author}{\bibfnamefont{A.~J.} \bibnamefont{Millis}},
  \bibinfo{journal}{\prb} \textbf{\bibinfo{volume}{69}},
  \bibinfo{pages}{245302} (\bibinfo{year}{2004}).

\bibitem[{\citenamefont{Egger and Gogolin}(2008)}]{Egger2008PRB}
\bibinfo{author}{\bibfnamefont{R.}~\bibnamefont{Egger}} \bibnamefont{and}
  \bibinfo{author}{\bibfnamefont{A.~O.} \bibnamefont{Gogolin}},
  \bibinfo{journal}{\prb} \textbf{\bibinfo{volume}{77}},
  \bibinfo{pages}{113405} (\bibinfo{year}{2008});
\bibitem[{\citenamefont{Entin-Wohlman et~al.}(2009)\citenamefont{Entin-Wohlman, Imry, and
  Aharony}}]{Wohlman2009PRB}
\bibinfo{author}{\bibfnamefont{O.} \bibnamefont{Entin-Wohlman}},
  \bibinfo{author}{\bibfnamefont{Y.}~\bibnamefont{Imry}}, \bibnamefont{and}
  \bibinfo{author}{\bibfnamefont{A.}~\bibnamefont{Aharony}},
  \bibinfo{journal}{\prb} \textbf{\bibinfo{volume}{80}},
  \bibinfo{pages}{035417} (\bibinfo{year}{2009}).

\bibitem[{\citenamefont{Wingreen et~al.}(1989)\citenamefont{Wingreen, Jacobsen,
  and Wilkins}}]{Wingreen1989PRB}
\bibinfo{author}{\bibfnamefont{N.~S.} \bibnamefont{Wingreen}},
  \bibinfo{author}{\bibfnamefont{K.~W.} \bibnamefont{Jacobsen}},
  \bibnamefont{and} \bibinfo{author}{\bibfnamefont{J.~W.}
  \bibnamefont{Wilkins}}, \bibinfo{journal}{\prb}
  \textbf{\bibinfo{volume}{40}}, \bibinfo{pages}{11834} (\bibinfo{year}{1989});
\bibitem[{\citenamefont{Flensberg}(2003)}]{Flensberg2003PRB}
\bibinfo{author}{\bibfnamefont{K.}~\bibnamefont{Flensberg}},
  \bibinfo{journal}{\prb} \textbf{\bibinfo{volume}{68}},
  \bibinfo{pages}{205323} (\bibinfo{year}{2003});
  \bibitem[{\citenamefont{Koch et~al.}(2004)\citenamefont{Koch, von Oppen, Oreg,
  and Sela}}]{Koch2004PRB}
\bibinfo{author}{\bibfnamefont{J.}~\bibnamefont{Koch}},
  \bibinfo{author}{\bibfnamefont{F.}~\bibnamefont{von Oppen}},
  \bibinfo{author}{\bibfnamefont{Y.}~\bibnamefont{Oreg}}, \bibnamefont{and}
  \bibinfo{author}{\bibfnamefont{E.}~\bibnamefont{Sela}},
  \bibinfo{journal}{\prb} \textbf{\bibinfo{volume}{70}},
  \bibinfo{pages}{195107} (\bibinfo{year}{2004}).


\bibitem[{\citenamefont{Kubala and K\"onig}(2002)}]{Kubala2002PRB}
\bibinfo{author}{\bibfnamefont{B.}~\bibnamefont{Kubala}} \bibnamefont{and}
  \bibinfo{author}{\bibfnamefont{J.}~\bibnamefont{K\"onig}},
  \bibinfo{journal}{\prb} \textbf{\bibinfo{volume}{65}},
  \bibinfo{pages}{245301} (\bibinfo{year}{2002});
\bibitem[{\citenamefont{Silva et~al.}(2002)\citenamefont{Silva, Oreg, and
  Gefen}}]{Silva2002PRB}
\bibinfo{author}{\bibfnamefont{A.}~\bibnamefont{Silva}},
  \bibinfo{author}{\bibfnamefont{Y.}~\bibnamefont{Oreg}}, \bibnamefont{and}
  \bibinfo{author}{\bibfnamefont{Y.}~\bibnamefont{Gefen}},
  \bibinfo{journal}{\prb} \textbf{\bibinfo{volume}{66}},
  \bibinfo{pages}{195316} (\bibinfo{year}{2002}).

\bibitem[{Note1()}]{Note1}\bibinfo{note}{Here, as in the remainder of this paper, we consider
  particle currents. The displacement-current contribution has to be treated
  separately \cite {Buttiker1993PLA}.}

\bibitem[{\citenamefont{Kubala and Marquardt}(2010)}]{Kubala2010PRB}
\bibinfo{author}{\bibfnamefont{B.}~\bibnamefont{Kubala}} \bibnamefont{and}
  \bibinfo{author}{\bibfnamefont{F.}~\bibnamefont{Marquardt}},
  \bibinfo{journal}{\prb} \textbf{\bibinfo{volume}{81}},
  \bibinfo{pages}{115319} (\bibinfo{year}{2010}).

\bibitem[{\citenamefont{Jauho et~al.}(1994)\citenamefont{Jauho, Wingreen, and
  Meir}}]{JauhoPRB94}
\bibinfo{author}{\bibfnamefont{A.~P.} \bibnamefont{Jauho}},
  \bibinfo{author}{\bibfnamefont{N.~S.} \bibnamefont{Wingreen}},
  \bibnamefont{and} \bibinfo{author}{\bibfnamefont{Y.}~\bibnamefont{Meir}},
  \bibinfo{journal}{\prb} \textbf{\bibinfo{volume}{50}}, \bibinfo{pages}{5528}
  (\bibinfo{year}{1994}).

\bibitem[{\citenamefont{Li and Su}(1996)}]{Li1996Quantum}
\bibinfo{author}{\bibfnamefont{X.~Q.} \bibnamefont{Li}} \bibnamefont{and}
  \bibinfo{author}{\bibfnamefont{Z.~B.} \bibnamefont{Su}},
  \bibinfo{journal}{\prb} \textbf{\bibinfo{volume}{54}}, \bibinfo{pages}{10807}
  (\bibinfo{year}{1996}).

\bibitem[{\citenamefont{Lehmann et~al.}(2004)\citenamefont{Lehmann, Kohler,
  May, and H\"{a}nggi}}]{Lehmann2004JCP}
\bibinfo{author}{\bibfnamefont{J.}~\bibnamefont{Lehmann}},
  \bibinfo{author}{\bibfnamefont{S.}~\bibnamefont{Kohler}},
  \bibinfo{author}{\bibfnamefont{V.}~\bibnamefont{May}}, \bibnamefont{and}
  \bibinfo{author}{\bibfnamefont{P.}~\bibnamefont{H\"{a}nggi}},
  \bibinfo{journal}{J. Chem. Phys.} \textbf{\bibinfo{volume}{121}},
  \bibinfo{pages}{2278} (\bibinfo{year}{2004}).

\bibitem[{\citenamefont{Costi et~al.}(1994)\citenamefont{Costi, Hewson, and
  Zlatic}}]{Costi1994}
\bibinfo{author}{\bibfnamefont{T.~A.} \bibnamefont{Costi}},
  \bibinfo{author}{\bibfnamefont{A.~C.} \bibnamefont{Hewson}},
  \bibnamefont{and} \bibinfo{author}{\bibfnamefont{V.}~\bibnamefont{Zlatic}},
  \bibinfo{journal}{J. Phys. Condens. Matter} \textbf{\bibinfo{volume}{6}},
  \bibinfo{pages}{2519} (\bibinfo{year}{1994}).

\bibitem[{\citenamefont{Hofstetter}(2000)}]{Hofstetter2000}
\bibinfo{author}{\bibfnamefont{W.}~\bibnamefont{Hofstetter}},
  \bibinfo{journal}{\prl} \textbf{\bibinfo{volume}{85}}, \bibinfo{pages}{1508}
  (\bibinfo{year}{2000}).

\bibitem[{\citenamefont{Weichselbaum and von Delft}(2007)}]{Weichselbaum2007}
\bibinfo{author}{\bibfnamefont{A.}~\bibnamefont{Weichselbaum}}
  \bibnamefont{and} \bibinfo{author}{\bibfnamefont{J.}~\bibnamefont{von
  Delft}}, \bibinfo{journal}{\prl} \textbf{\bibinfo{volume}{99}},
  \bibinfo{eid}{076402} (\bibinfo{year}{2007}).

\bibitem[{\citenamefont{Bulla et~al.}(2008)\citenamefont{Bulla, Costi, and
  Pruschke}}]{Bulla2008}
\bibinfo{author}{\bibfnamefont{R.}~\bibnamefont{Bulla}},
  \bibinfo{author}{\bibfnamefont{T.~A.} \bibnamefont{Costi}}, \bibnamefont{and}
  \bibinfo{author}{\bibfnamefont{T.}~\bibnamefont{Pruschke}},
  \bibinfo{journal}{\rmp} \textbf{\bibinfo{volume}{80}}, \bibinfo{eid}{395}
  (\bibinfo{year}{2008}).

\bibitem[{\citenamefont{Bulla et~al.}(2003)\citenamefont{Bulla, Tong, and
  Vojta}}]{Bulla2003}
\bibinfo{author}{\bibfnamefont{R.}~\bibnamefont{Bulla}},
  \bibinfo{author}{\bibfnamefont{N.-H.} \bibnamefont{Tong}}, \bibnamefont{and}
  \bibinfo{author}{\bibfnamefont{M.}~\bibnamefont{Vojta}},
  \bibinfo{journal}{\prl} \textbf{\bibinfo{volume}{91}},
  \bibinfo{pages}{170601} (\bibinfo{year}{2003}).

\bibitem[{\citenamefont{Sindel et~al.}(2005)\citenamefont{Sindel, Hofstetter,
  von Delft, and Kindermann}}]{Sindel2005PRL}
\bibinfo{author}{\bibfnamefont{M.}~\bibnamefont{Sindel}},
  \bibinfo{author}{\bibfnamefont{W.}~\bibnamefont{Hofstetter}},
  \bibinfo{author}{\bibfnamefont{J.}~\bibnamefont{von Delft}},
  \bibnamefont{and}
  \bibinfo{author}{\bibfnamefont{M.}~\bibnamefont{Kindermann}},
  \bibinfo{journal}{\prl} \textbf{\bibinfo{volume}{94}},
  \bibinfo{pages}{196602} (\bibinfo{year}{2005});
\bibitem[{\citenamefont{Moca et~al.}(2010)\citenamefont{Moca, Weymann, and
  Zar\'{a}nd}}]{Moca2010PRB}
\bibinfo{author}{\bibfnamefont{C.~P.} \bibnamefont{Moca}},
  \bibinfo{author}{\bibfnamefont{I.}~\bibnamefont{Weymann}}, \bibnamefont{and}
  \bibinfo{author}{\bibfnamefont{G.}~\bibnamefont{Zar\'{a}nd}},
  \bibinfo{journal}{\prb} \textbf{\bibinfo{volume}{81}},
  \bibinfo{pages}{241305} (\bibinfo{year}{2010}).

\bibitem[{\citenamefont{Hewson and Meyer}(2002)}]{Hewson2002}
\bibinfo{author}{\bibfnamefont{A.~C.} \bibnamefont{Hewson}} \bibnamefont{and}
  \bibinfo{author}{\bibfnamefont{D.}~\bibnamefont{Meyer}}, \bibinfo{journal}{J.
  Phys. Condens. Matter} \textbf{\bibinfo{volume}{14}}, \bibinfo{pages}{427}
  (\bibinfo{year}{2002}).

\bibitem[{\citenamefont{Cornaglia et~al.}(2004)\citenamefont{Cornaglia, Ness,
  and Grempel}}]{Cornaglia2004}
\bibinfo{author}{\bibfnamefont{P.~S.} \bibnamefont{Cornaglia}},
  \bibinfo{author}{\bibfnamefont{H.}~\bibnamefont{Ness}}, \bibnamefont{and}
  \bibinfo{author}{\bibfnamefont{D.~R.} \bibnamefont{Grempel}},
  \bibinfo{journal}{Phys. Rev. Lett.} \textbf{\bibinfo{volume}{93}},
  \bibinfo{pages}{147201} (\bibinfo{year}{2004});
\bibitem[{\citenamefont{Cornaglia and Grempel}(2005)}]{Cornaglia2005}
\bibinfo{author}{\bibfnamefont{P.~S.} \bibnamefont{Cornaglia}}
  \bibnamefont{and} \bibinfo{author}{\bibfnamefont{D.~R.}
  \bibnamefont{Grempel}}, \bibinfo{journal}{Phys. Rev. B}
  \textbf{\bibinfo{volume}{71}}, \bibinfo{pages}{245326}
  (\bibinfo{year}{2005}).

\bibitem[{\citenamefont{Yoshida et~al.}(1990)\citenamefont{Yoshida, Whitaker,
  and Oliveira}}]{Yoshida1990}
\bibinfo{author}{\bibfnamefont{M.}~\bibnamefont{Yoshida}},
  \bibinfo{author}{\bibfnamefont{M.~A.} \bibnamefont{Whitaker}},
  \bibnamefont{and} \bibinfo{author}{\bibfnamefont{L.~N.}
  \bibnamefont{Oliveira}}, \bibinfo{journal}{\prb}
  \textbf{\bibinfo{volume}{41}}, \bibinfo{pages}{9403} (\bibinfo{year}{1990});
\bibitem[{\citenamefont{Zitko and Pruschke}(2009)}]{Zitko2009}
\bibinfo{author}{\bibfnamefont{R.}~\bibnamefont{Zitko}} \bibnamefont{and}
  \bibinfo{author}{\bibfnamefont{T.}~\bibnamefont{Pruschke}},
  \bibinfo{journal}{\prb} \textbf{\bibinfo{volume}{79}}, \bibinfo{eid}{085106}
  (\bibinfo{year}{2009}).

\end{thebibliography}

\end{document}